\documentstyle[11pt,pictex]{article}

\newcommand{\beq}{\begin{equation}}
\newcommand{\eeq}{\end{equation}}
\newcommand{\beqa}{\begin{eqnarray}}
\newcommand{\eeqa}{\end{eqnarray}}
\newcommand{\nn}{\nonumber}
\newcommand{\eq}[1]{(\ref{#1})}

\newcommand{\ra}{\rightarrow}

\newcommand{\tr}{{\rm Tr}}
\newcommand{\ogamma}{\Gamma{\mskip-12mu\raise1.5pt
\vbox{\kern7pt\hbox{${\tt o}$}}\mskip1mu}}

\newcommand{\NP}[1]{ {\it Nucl.{}~Phys.} {\bf #1}}
\newcommand{\PL}[1]{ {\it Phys.{}~Lett.} {\bf #1}}

\newcommand{\PR}[1]{ {\it Phys.{}~Rev.} {\bf #1}}

\newcommand{\AP}[1]{ {\it Ann.{}~Phys.} {\bf #1}}

\newcommand{\WS}[1]{ (World Scientific Publishing, #1)}

\renewcommand{\theequation}{\thesection.\arabic{equation}}
\begin{document}
\topmargin 0pt
\oddsidemargin 1mm
\begin{titlepage}
%%%\rightline{ }
\begin{flushright}
HD-THEP-98-8\\
hep-th/9802127\\
%manuscript
%draft v4, 10/2/98
\end{flushright}
\setcounter{page}{0}

\vspace{15mm}
\begin{center}
{\Large Exact Combinatorics of Bern-Kosower-type 
Amplitudes for Two-Loop $\Phi^3$ Theory} 
\vspace{20mm}

{\large Haru-Tada Sato{}~\footnote{h.sato@thphys.uni-heidelberg.de}} 
{\sc and }  
{\large Michael G. Schmidt   
\footnote{m.g.schmidt@thphys.uni-heidelberg.de}}\\
{\em Institut f{\"u}r Theoretische Physik, Universit{\"a}t Heidelberg\\
     Philosophenweg 16, D-69120 Heidelberg, Germany}\\
\end{center}
\vspace{7mm}

\begin{abstract}
Counting the contribution rate of a world-line formula to 
Feynman diagrams in $\phi^3$ theory, 
we explain the idea how to determine precise combinatorics of 
Bern-Kosower-like amplitudes derived from a bosonic string theory 
for $N$-point two-loop Feynman amplitudes. In this connection we 
also present a method to derive simple and compact world-line 
forms for the effective action. 
\end{abstract}

\vspace{1cm}
\begin{flushleft}
\end{flushleft}

\end{titlepage}
\newpage
\renewcommand{\theequation}{\thesection.\arabic{equation}}
\renewcommand{\thefootnote}{\arabic{footnote}}
%%%%%%%%%%%%%%%%%%%%%%%%%%%%%%%%%%%%%%%%%%%%%%%%%%%%%%%%%%%%%%%%%%%%%%%%%%%%
%                                                                          %
%              Sect.1      Introduction                                    %
%                                                                          %
%%%%%%%%%%%%%%%%%%%%%%%%%%%%%%%%%%%%%%%%%%%%%%%%%%%%%%%%%%%%%%%%%%%%%%%%%%%%
\section{Introduction}
\setcounter{equation}{0}
\indent

Field theory can be understood as a singular limit of string theory, 
and the relation between both theories has been investigated intensively 
for the purpose of obtaining field theory scattering amplitudes in a 
remarkably simple way~\cite{BK}-\cite{paolo}. String theory organizes 
scattering amplitudes in a compact form because of conformal symmetry 
on the world-sheet, and field theory would inherit this useful feature, by 
which summation of Feynman diagrams is already installed. In particular, 
Bern and Kosower derived a set of simple rules for one-loop gluon 
scattering amplitudes through analyzing the field theory limit of a 
heterotic string theory~\cite{BK}. Later on, it was realized that these 
rules can also be derived directly in the world-line approach to quantum 
field theory~\cite{St} and that effective actions can be evaluated most 
conveniently with this method \cite{EA}. 
There are also applications to gravity \cite{gra} and 
super Yang-Mills theories \cite{sYM}. 

It is also interesting to find a multi-loop generalization of Bern-Kosower 
rules, and various steps toward this direction were made in recent 
years~\cite{SSphi}-\cite{xxx}. The most well-understood theory is the 
$\phi^3$ theory~\cite{SSphi}-\cite{RS2}. Universal expressions 
(master formulae) for proper $N$-point functions were derived from 
field theory (world-line approach) \cite{SSphi,HTS} and bosonic 
string theories~\cite{RS1}-\cite{RS2}; the correspondence of corners of moduli 
to Feynman diagrams~\cite{RS1,paolo1}, the field theory limits of 
world-sheet Green function~\cite{RS1}, and the determinant factor for 
moduli integrals~\cite{RS2} were examined in detail. The results in 
both cases coincide with each other up to a combinatorial problem, 
which should finally be solved to construct the complete $N$-point 
functions. The master formulae contain various Feynman integrals 
labeled by a set of integers which represent the numbers of external 
legs inserted in internal lines as basic parts of world-line 
parametrization, and the problem is how to combine the formulas of 
different set of integers for fixed $N$ in order to make up the 
desired result. 

In this paper we solve this problem in the two-loop case for $\phi^3$ 
theory, for the purpose of a suggestion to multi-loop generalization. 
The basic idea to obtain exact $N$-point proper functions 
from the master formula is the following. {}~From either string or 
world-line theories, we can derive the $N$-point master formula, 
which takes, for example, the following form in the one-loop case 
\beq
\Gamma_M^{1-loop} = (-g)^N\int_0^\infty dT(4\pi T)^{-D/2}\int_M
\prod_{n=1}^{N-1} d\tau_n \exp[ \sum_{i<j}^N p_i\cdot p_j G_B(\tau_i,\tau_j)]\Bigr|_{\tau_N=T} \ ,        \label{oneloop}
\eeq
where $G_B$ is the one-loop bosonic Green function 
\beq
G_B(\tau_1,\tau_2)=|\tau_1-\tau_2|-{(\tau_1-\tau_2)^2\over T}\ ,
\eeq
and $M=\{0\leq\tau_n\leq T; n=1,\dots,N-1\}$. 
(Here, we have removed the pre-factor $1/2$ on RHS in \eq{oneloop}  
as a matter of convenience for generic arguments). Let us divide the 
whole region $M$ into $(N-1)!$ sub-regions according to the orderings 
of $\tau_n$, $n=1,\dots,N-1$. One can re-arrange a $N$-point Feynman 
integral in the same form as \eq{oneloop} restricted to a certain 
sub-region of the master region $M$. In this way the master formula 
comprises the $N$-point Feynman integral and its topologically 
independent diagrams {\it at least once}. The essential quantity which 
we are going to discuss in this paper is the ratio ({\it covering 
multiplicity}) between the number of independent Feynman diagrams 
and the number of corresponding sub-regions. In order to extract 
each Feynman diagram's contribution exactly once, we have to 
determine this ratio. In the one-loop case, this ratio is just 2, 
and we have only to put the inverse of this number in front 
of the master formula \eq{oneloop}. 

In the two-loop case, we, in principle, follow the same procedure. 
However we need more careful treatment than in the one-loop case. 
We have complications for the following reasons: 
(i) The manner of dividing a master integration region depends on 
how to parametrize the 1PI vacuum diagram part. In other words, it 
depends on the definition of $M$, in which regions each $\tau_n$ 
runs. In this way, the numbers of sub-regions and their diagram 
multiplicities also depend on the parametrizations. We have no 
general idea which choice of parametrization is the most convenient 
one. (ii) We also have to split $N$, the number of external legs, 
into a sum of integers like $N=N_1+N_2+N_3$ or $N=N'+N_3$ in 
accordance with the choice of parametrizations. This delivers 
all Feynman diagrams into a certain number of diagram classes, 
and each diagram class possesses its own number of Feynman 
diagrams, sub-regions, and thus its own covering multiplicity. 
(iii) Since covering multiplicities are different class by class, 
the pre-factor corrections for master formulas are no longer the 
simple inverse of the multiplicity. Taking account of these covering 
multiplicities for each class, we gather 
them all to make a correct proper $N$-point function expressed in 
terms of a combination of master formulas. 
In the world-line formalism, the effective action is a starting 
point, and in fact the 1PI $N$-point functions are nothing but 
the Fourier transforms (plane wave expansions) of effective action. 
This viewpoint is certainly helpful to understand a combinatorial 
structure of a master formula, and we will also see that the 
effective action can be written in the world-line language most 
conveniently.

This paper is organized as follows.
In sect. 2, we explain our notational settings. 
In sect. 3, we propose a method how to determine the number of 
topologically independent Feynman diagrams. This is based on 
the two-loop effective action extracted from an auxiliary field 
formalism combined with the background field method. 
Then we analyze two parametrizations case by case 
in the order of simplicity. The first one is a loop type 
parametrization which regards $N'$ legs are on a circle 
({\it fundamental loop}) and $N_3$ legs are on the remaining 
internal line (which we simply call {\it middle line}). In sect. 4, 
we confine to the $N_3=0$ cases. These cases are useful to 
determine the precise normalization of the photon scattering master 
formula in QED. In sect. 5, we discuss the symmetric parametrization 
which regards $N_a$, $a=1,2,3$ legs are on the three internal 
lines. In sect. 6, we also discuss a few sample cases for 
$N_3\not=0$ in the loop type parametrization. We shall not give 
a general prescription in these cases. Some technical details are 
available in the appendices: In Appendix A we explain how the 
combinatorics of the symmetric master formula emerges from the 
effective action in $\phi^3$-theory, and in Appendix B we remark on 
the translational invariance along a super world-line fundamental 
loop, on which the invariance is necessary to connect field theory 
limits of string amplitudes with world-line formulations. 

%%%%%%%%%%%%%%%%%%%%%%%%%%%%%%%%%%%%%%%%%%%%%%%%%%%%%%%%%%%%%%%%%%%%%%%%%%
%            Sect. 2
%%%%%%%%%%%%%%%%%%%%%%%%%%%%%%%%%%%%%%%%%%%%%%%%%%%%%%%%%%%%%%%%%%%%%%%%%%
\section{Notations for Feynman amplitudes and world-line formulae}
\setcounter{equation}{0}
\indent

Let us classify all two-loop (proper) Feynman diagrams in accord with 
$N_a$, $a=1,2,3$, the numbers of external legs which are inserted in 
each of the three internal lines labeled by $T_a$, $a=1,2,3$. Ignoring 
the ordering of external legs, we define a representative symbol diagram 
$F_i$ ($N_1\geq N_2\geq N_3$), where the label $i$ (which we refer to 
as {\it class}) may be chosen as $(N_1,N_2,N_3)$. This diagram 
$F_i$ is actually an element of its diagram set ${\cal T}_i$ which 
consists of all topologically inequivalent diagrams $d_n$, 
$n=1,\dots,n^T_i$, obtained by shuffling all leg orderings of $F_i$. 
The $N$-point Feynman amplitudes $\Gamma_N$ are obtained by the 
following three steps: (i)  For a diagram $d_n$ of a class $i$, write 
down the Feynman amplitude $\Gamma[d_n]$ including its symmetry 
factor $S_i$, which is common for all $d_n$ in the set ${\cal T}_i$. 
(ii) Sum up all $\Gamma[d_n]$ over the set ${\cal T}_i$, i.e. 
\beq
{\bar\Gamma}_{F_i} = \sum_{d_n\in {\cal T}_i} \Gamma[d_n]\ .
\eeq
Obviously, we simply have 
$\Gamma_{F_i}\equiv{\bar\Gamma}_{F_i}=\Gamma[d_1]$, 
if a class $i$ is composed of only one diagram, $n^T_i =1$. 
(iii) Finally sum up the ${\bar\Gamma}_{F_i}$ of all classes to obtain
\beq
      \Gamma_N = \sum_i {\bar\Gamma}_{F_i}.
\eeq
If we exclude from ${\cal T}_i$ the diagrams where external legs 
are inserted in the {\it middle line} propagator, in our notation 
$\Gamma$ will be denoted by $\ogamma$
\beq
{\bar\ogamma}_{F_i}\ , \qquad \ogamma_N \ , \quad \mbox{etc}.
\eeq
The symmetry factors of the theory are generally given by (see \cite{muta}) 
\beq
S = \Bigl[ p_1 (2!)^{p_2} (3!)^{p_3} \Bigr]^{-1}\ ,\label{Sfactor}
\eeq
where $p_n$, $n$ = 2,3, are the numbers of vertex pairs connected 
directly by $n$ lines, and $p_1$ is the number of vertex permutations 
which leave the diagram unchanged with external legs held fixed. (For 
a tadpole part, take $p_2=1$). 

In the world-line formalism, we have two ways of parametrizing the 
1PI vacuum diagram~\cite{SSphi}. The first one is the symmetric 
parametrization, in which we deal with each $T_i$ on equal ground and 
$N_a$ external legs are inserted in each line $T_a$. The other is the 
loop-type parametrization, where two lines of $T_a$ are combined to 
form a {\it fundamental loop} parameter $T$ ($=T_1+T_2$) and $N'=N_1+N_2$ 
legs are inserted in the fundamental loop. (Obviously we do not restrict 
to $N_1\leq N_2\leq N_3$ at this stage.) Thus we have the following 
two representations of a master formula for the two-loop $N$-point 
amplitudes (see \cite{RS1} for notational details). The loop type 
master formula is given by~\cite{SSphi,HTS,RS1}  
\beqa
&&\Gamma_M^{(N',N_3)} =
\Gamma_M^{(N',N_3)}(p_1,\cdots,p_{N'}|p_1^{(3)},\cdots,p_{N_3}^{(3)})\nn\\
&=&{(-g)^{N+2}\over(4\pi)^D}\cdot
\int_0^{\infty}dT d{\bar T}e^{-m^2(T+{\bar T})}
 \int_M \prod_{n=1}^{N} d\tau_n  d\tau_\beta
[T{\bar T}+TG_B(\tau_\alpha,\tau_\beta)]^{-D/2}  \cdot \label{masterL} \\
&\times& \left.\exp[{1\over2}\sum_{jk}^{N'} p_j p_k 
G^{(1)}_{11}(\tau_j,\tau_k)+{1\over2}\sum_{jk}^{N_3} p^{(3)}_j p^{(3)}_k 
G^{(1)}_{33}(\tau^{(3)}_j,\tau^{(3)}_k)
+\sum_j^{N'}\sum_k^{N_3} p_j p^{(3)}_k 
G^{(1)}_{13}(\tau_j, \tau^{(3)}_k) ]\right|_{\tau_\alpha=0} \nn
\eeqa  
where ${\bar T}\equiv T_3$. The normalization $(4\pi)^{-D}$ also 
follows from string theory~\cite{RS2}. 
Introducing obvious cyclic notations such as $p^{(4)}=p^{(1)}$ 
etc., the symmetric master formula is given by~\cite{SSphi,HTS} 
\beqa
&&\Gamma_M^{(N_1,N_2,N_3)} =
\Gamma_M^{(N_1,N_2,N_3)}(p_1^{(1)},\cdots,p_{N_1}^{(1)}| 
p_1^{(2)},\cdots,p_{N_2}^{(2)}|p_1^{(3)},\cdots,p_{N_3}^{(3)})\nn\\
&=& {(-g)^{N+2}\over(4\pi)^D}\cdot
\prod_{a=1}^3 \int_0^{\infty}dT_a e^{-m^2 T_a} \cdot 
(T_1T_2+T_2T_3+T_3T_1)^{-D/2} 
 \int_M \prod_{n=1}^{N} d\tau_n  \label{masterS} \\
&\times& \exp[{1\over2}\sum_{a=1}^3 \sum_{j,k}^{N_a} p^{(a)}_j p^{(a)}_k 
       G^{\rm sym}_{aa}(\tau^{(a)}_j,\tau^{(a)}_k)
   +\sum_{a=1}^3\sum_j^{N_a}\sum_k^{N_{a+1}} p^{(a)}_j p^{(a+1)}_k 
   G^{\rm sym}_{aa+1}(\tau^{(a)}_j,\tau^{(a+1)}_k) ]\ . \nn
\eeqa
In both formulae, the subscript $M$ stands for the full 
integration regions of all $\tau$-parameters 
\beq
M = \left\{ \begin{array}{ll}
\{0\leq\tau_n\leq T,\,0\leq\tau^{(3)}_m\leq T_3\,|\,
n=1,\dots,N',\beta\,;\, m=1,\dots,N_3\,\}
&\quad\mbox{for}\quad \eq{masterL}\\
\{0\leq\tau^{(a)}_{n_a}\leq T^{(a)}\,|\, 
n_a=1,\dots,N_a\,;\, a=1,2,3\,\}
&\quad\mbox{for}\quad \eq{masterS}\ . 
\end{array} \right.   
\eeq
Since the splitting of the integrations over $M$ 
depends on the choice of parametrization, these two master formulae 
possess different diagram contents, to be more precise, different 
covering multiplicities for Feynman diagrams. It means that 
\eq{masterL} and \eq{masterS} do not by themselves coincide with 
each other. However there is a simple relation between them 
focusing on a divided (ordered $\tau$-) integration region in each 
formula~\cite{HTS}
\beq
\Gamma[d_n] = S\Gamma_{D_k}^{(N'N_3)} = 
S\Gamma_{D_{k'}}^{(N_1N_2N_3)}\ ,
\eeq
where $D_k$ and $D_{k'}$ express regions obtained by splitting 
the respective master regions $M$. 

To obtain the $N$-point function $\Gamma_N$, we certainly have to 
use the master formula to sum up over all possible sets of 
$(N_1,N_2,N_3)$ or of $(N',N_3)$ with some constant weights in each 
case. The main question is which factors should appear in front of 
these master formulae in order to really obtain $\Gamma_N$ 
or $\ogamma_N$. 

%%%%%%%%%%%%%%%%%%  Sect.3   BGF method %%%%%%%%%%%%%%%%%%%%%%%%%%% 
\section{Background field plus auxiliary field method}
\setcounter{equation}{0}
%%%%%%%%%%%%%%%%%%%%%%%%%%%%%%%%%%%%%%%%%%%%%%%%%%%%%%%%%%%%%%%%%%%%
\indent

In order to obtain the covering multiplicities, it is necessary to 
know $n^T_i$ the number of topologies for each diagram class $F_i$. 
In this section~\footnote{We appreciate a contribution of M. Reuter 
to this section~\cite{xx}.}, we present a method to determine the 
value of $n^T_i$. In principle, this information is contained in 
the generating functional. Let us consider the generating functional 
for the Euclidean Lagrangian 
\beq
{\cal L} = {1\over2}(\partial\phi\partial\phi+m^2\phi^2) 
          + {g\over3!}\phi^3
\eeq
in the background field method decomposing 
$\phi=\varphi+{\bar\phi}$, where $\varphi$ is a quantum field while 
${\bar\phi}$ is a classical field. This produces the 
generating functional  
\beq
Z[{\bar\phi}] =Z_0\int{\cal D}\varphi
\exp\Bigl[-\int\Bigl\{ 
{1\over2}\varphi(\Delta^{-1}+g{\bar\phi})\varphi 
+ {g\over3!}\varphi^3\Bigr\}
d^D x\Bigr] \ ,              \label{generate}
\eeq
where $\Delta^{-1}$ is the free inverse propagator $\Delta^{-1}= 
-\partial^2+m^2$, and $Z_0$ consists of classical (tree) 
terms
\beq
Z_0 = \exp\Bigl[\int d^Dx\Bigl\{
-{1\over2}{\bar\phi}\Delta^{-1}{\bar\phi}-{g\over3!}{\bar\phi}^3
\Bigr\}\,\Bigr]\,. 
\eeq
In order to perform the $\varphi$ integration, 
we further introduce the auxiliary field $B$ which represents 
$\varphi^2$ by insertion of a delta function
\beq
\delta(B-\varphi^2)= \int{\cal D}\alpha\exp\Bigl[i
\int \alpha(B-\varphi^2) d^Dx\,\Bigr]\ . \label{unit}
\eeq
The quantum part of the Lagrangian in \eq{generate} then reads   
\beq
{\cal L}^{new}={1\over2}\varphi(\Delta^{-1}+i2\alpha)\varphi 
+\lambda B\varphi +3\lambda{\bar\phi}B 
-iB\alpha \ ,
\eeq
where 
\beq
\lambda={1\over3!}g \ .
\eeq
The new expression for $Z[{\bar\phi}]$ is 
\beqa
Z[{\bar\phi}] &=& Z_0 \int{\cal D}\alpha{\cal D}B
\mbox{Det}^{-1/2}(\Delta^{-1}+i2\alpha) \nn\\
&&\exp[\,{\lambda^2\over2} B(\Delta^{-1}+i2\alpha)^{-1}B 
- 3\lambda{\bar\phi}B + iB\alpha\,] \ ,
\eeqa
where an integration over space-time is understood in the exponent. 
Applying the following formula for a function of $\alpha$
\beq
\int{\cal D}\alpha{\cal D}Bf(i\alpha)e^{iB\alpha}=
\int{\cal D}\alpha{\cal D}Bf({\partial\over\partial B})e^{iB\alpha}
=\int{\cal D}Bf({\partial\over\partial B})\delta(B)\ ,
\label{deformulaB}
\eeq 
we rewrite
\beq
Z[{\bar\phi}] =Z_0 \cdot\exp\Bigl[ -{1\over2}
\mbox{Tr}\ln(\Delta^{-1}-2\partial_B)\Bigr]
\exp[\,{\lambda^2\over2}
B(\Delta^{-1}-2\partial_B)^{-1}B \nn\\
 - 3\lambda{\bar\phi}B \,]\Bigr|_{B=0}\ . \label{zphibarfinal}
\eeq
Note that the sign of $\partial_B$ is reversed because of 
a partial integration for the delta function $\delta(B)$. 

Here we put a remark on an alternative calculation. 
We can also apply the formula similar to \eq{deformulaB} 
by exchanging $\alpha$ and $B$. In this case the last term in the 
exponent of \eq{zphibarfinal} becomes 
$-i3\lambda{\bar\phi}\partial_\alpha$, which provides a 
simultaneous translation of all $\alpha$'s, and then 
\beq
Z[{\bar\phi}] =Z_0 \cdot\exp\Bigl[ -{1\over2}
\mbox{Tr}\ln(\Delta^{-1}+g{\bar\phi}+2i\alpha)\Bigr]
\exp[\,-{\lambda^2\over2}\partial_\alpha
(\Delta^{-1}+g{\bar\phi}+2i\alpha)^{-1}\partial_\alpha 
\,]\Bigr|_{\alpha=0}\ . \label{zphialpha}
\eeq
As to two-loop contributions, we may pick up second order terms 
in $\partial_\alpha$, which act on the $\alpha$ field in the $\tr\ln$ 
loop term as well. Applying a path integral representation for 
(open/closed) propagator (q.v. \eq{pathprop}), we have 
\beq
Z^{2-loop}=Z_0\lambda^2(2I_1+I_2)
\eeq
with
\beqa
I_1&=&I_1[0\leq\tau_1,\tau_2\leq S]
= \int_0^\infty dS\int_0^S d\tau_1\int_0^S d\tau_2 
 \int_{\scriptstyle y(0)=y(\tau_1) 
      \atop\scriptstyle y(S)=y(\tau_2)}{\cal D}y
e^{-A(y,S)} \\
I_2&=&\int_0^\infty{dT\over T}\oint{\cal D}xe^{-A(x,T)}
\int_0^Td\tau_1\int_0^Td\tau_2\int_0^\infty d{\bar T} 
\int_{\scriptstyle y(0)=x(\tau_1) 
   \atop\scriptstyle y({\bar T})=x(\tau_2)}{\cal D}y
e^{-A(y,{\bar T})}
\eeqa
where 
\beq
A(x,T) =\int_0^T({1\over4}{\dot x}^2(\tau)+g{\bar\phi}(x))d\tau\ .
\eeq
Note that $I_2$ and $I_1$ correspond to (a) and (b) in Fig. 2 
respectively. An interesting observation is that $I_1$ can be 
reduced to $I_2$ if we confine to the 1PI parts. Restricting 
to the region $0\leq\tau_2\leq\tau_1$ (outer region corresponds 
to 1PR parts) and changing variables in integrations 
$\tau_1=T$, $S=T+{\bar T}$, we find out 
\beq
I_1^{1PI}=I_1[0\leq\tau_2\leq\tau_1\leq S] =
\int_0^\infty d{\bar T}\int_0^\infty dT\int_0^T d\tau_2
\oint{\cal D}x \int_{\scriptstyle y(0)=x(T) 
      \atop\scriptstyle y({\bar T})=x(\tau_2)}{\cal D}y
e^{-A(x,T)-A(y,{\bar T})}\ .
\eeq
{}~Fixing $\tau_1=T$ in $I_2$ (q.v. translational invariance 
in \eq{masterL}), we see $I_1^{1PI}=I_2$ and therefore 
\beq
\Gamma^{2-loop}={g^2\over2\cdot3!}I_2 
={g^2\over2\cdot3!}I_1^{1PI} \ .
\eeq
If one further changes integration variables as $T_1=\tau_2$, 
$T_2=\tau_1-\tau_2$, $T_3=S-\tau_1$ ($S=T_1+T_2+T_3$), 
one can reproduce the symmetric three-propagator expression 
shown in Appendix A. 

Now let us go back to the main purpose mentioned in the heading of 
the section. Eq.\eq{zphibarfinal} can be expanded 
in a perturbative expansion form
\beq
Z[{\bar\phi}]= \sum_{n=0}^\infty (-g)^n z_n[{\bar\phi}]\ ,
\eeq
and the function $z_n[{\bar\phi}]$ possesses the following structure: 
Let $F_i$ be a representative symbol diagram of order $g^n$ without 
having any tree vertex, and $F_i[{\bar\phi}]$ be the coordinate space 
representation of $F_i$ in terms of $\Delta$ and $\Delta{\bar\phi}$. 
Then $z_n[{\bar\phi}]$ is given by a sum of $F_i[{\bar\phi}]$ with 
certain weights $w_i$;
\beq
z_n [{\bar\phi}] =  \sum_i w_i F_i[{\bar\phi}] \ . \label{zn}
\eeq
To obtain these weights, one may just calculate $Z[{\bar\phi}]$ 
term by term at each order of $g$. For example $z_1$ and $z_2$ are 
given by 
\beq
z_1[{\bar\phi}]={1\over2}\int d^Dx_1 
              \Delta_{11}(\Delta{\bar\phi})_{11}\ ,
\eeq
\beqa
z_2[{\bar\phi}]={1\over2(3!)^2}\int d^Dx_1d^Dx_2&\Bigl[&6\Delta^3_{12}
+9\Delta_{11}\Delta_{12}\Delta_{22}+9\Delta_{11}\Delta_{22}
(\Delta{\bar\phi})_{11}(\Delta{\bar\phi})_{22} \nn\\
&&+18\Delta^2_{12}(\Delta{\bar\phi})_{11}(\Delta{\bar\phi})_{22}
\,\Bigr]\ ,
\eeqa
where
\beq
\Delta_{ij}=\Delta(x_i,x_j)=(-\partial^2+m^2)^{-1}_{ij}
=\int{d^Dk\over(2\pi)^D}{e^{-ik(x_i-x_j)}\over m^2+k^2}\ ,
\eeq
\beq
(\Delta{\bar\phi})_{ij}=\int d^Dy_j\Delta(x_i,y_j){\bar\phi}(y_j)\ .
\eeq

The general rules how to determine the coefficients 
$w_i$ are the following: (i) Consider the symbol diagram 
$F_i$ possessing $N$ external legs, and compose the same diagram 
from sewing the diagram parts depicted in Fig. 1. To sew the parts, 
one must insert some $\partial_B$'s on a $<BB>$ line or on a loop 
as many as $n$ the order of $g$ (Do not insert on a $<B{\bar\phi}>$ 
line). Then join the $\partial_B$-crosses and the $B$-dots.  
(ii) Assign the following numerical factors in each sewn diagram
\beq
\left.\begin{array}{ll}
2   &\quad\mbox{for}\quad   \partial_B    \\
({1\over2}\lambda^2)^n/n! &\quad\mbox{for $n$ propagators}\quad <BB>^n  \\
(3\lambda)^n/n! &\quad\mbox{for $n$ external legs}\quad <B{\bar\phi}>^n \\
%(3\lambda^2)^n/n! &\quad\mbox{for $n$ 1PR vertices}\quad (B{\bar\phi}^2)^n \\
(2^L L!\prod_{i=1}^L n_i)^{-1} &\quad\mbox{for}\quad \mbox{$L$ loops 
from Trln part} \\
\end{array}\right. 
\eeq
where $n_i$ is the number of vertices on the $i$-th loop. 
(iii) Finally summing up the factors of all possible sewing diagrams 
for the $F_i$, we obtain the coefficient 
$w_i$ for $F_i$ (setting the coupling $\lambda\ra 1/3!$). 
An example, in the case of $F_i[{\bar\phi}]=\Delta_{12}^3$, is 
shown in Fig. 2, where 3 sewing possibilities exist. The two 
possibilities are represented by the diagram (a) which also 
includes an upside-down attachment of the $<BB>$-line to the loop, 
and the remaining one possibility is the diagram (b). 

%
% Pictex
\vspace{-5mm}
\begin{minipage}[t]{13cm} 
\font\thinlinefont=cmr5
\begingroup\makeatletter\ifx\SetFigFont\undefined
% extract first six characters in \fmtname
\def\x#1#2#3#4#5#6#7\relax{\def\x{#1#2#3#4#5#6}}%
\expandafter\x\fmtname xxxxxx\relax \def\y{splain}%
\ifx\x\y   % LaTeX or SliTeX?
\gdef\SetFigFont#1#2#3{%
  \ifnum #1<17\tiny\else \ifnum #1<20\small\else
  \ifnum #1<24\normalsize\else \ifnum #1<29\large\else
  \ifnum #1<34\Large\else \ifnum #1<41\LARGE\else
     \huge\fi\fi\fi\fi\fi\fi
  \csname #3\endcsname}%
\else
\gdef\SetFigFont#1#2#3{\begingroup
  \count@#1\relax \ifnum 25<\count@\count@25\fi
  \def\x{\endgroup\@setsize\SetFigFont{#2pt}}%
  \expandafter\x
    \csname \romannumeral\the\count@ pt\expandafter\endcsname
    \csname @\romannumeral\the\count@ pt\endcsname
  \csname #3\endcsname}%
\fi
\fi\endgroup
\mbox{\beginpicture
\setcoordinatesystem units <0.50000cm,0.50000cm>
\unitlength=0.50000cm
\linethickness=1pt
\setplotsymbol ({\makebox(0,0)[l]{\tencirc\symbol{'160}}})
\setshadesymbol ({\thinlinefont .})
\setlinear
%
% Fig ELLIPSE  a ---- BB end
%
\linethickness=7pt
\setplotsymbol ({\thinlinefont .})
%\setplotsymbol ({\makebox(0,0)[l]{\tencirc\symbol{'166}}})
\put{\makebox(0,0)[l]{\circle*{ 0.449}}} at 23.546 16.717
%
% Fig ELLIPSE  b ---- BB end
%
\linethickness=7pt
\setplotsymbol ({\makebox(0,0)[l]{\tencirc\symbol{'166}}})
\put{\makebox(0,0)[l]{\circle*{ 0.449}}} at 18.008 16.669
%
% Fig ELLIPSE --- phi B line end
%
\linethickness=7pt
\setplotsymbol ({\makebox(0,0)[l]{\tencirc\symbol{'166}}})
\put{\makebox(0,0)[l]{\circle*{ 0.449}}} at 14.064 16.669
%
% Fig ELLIPSE  ---- big circle
%
\linethickness=0.500pt
\setplotsymbol ({\thinlinefont .})
\ellipticalarc axes ratio  3.493:3.493  360 degrees 
%	from 33.574 16.669 center at 29.981 16.669
	from 33.174 16.669 center at 29.981 16.669

%
% Fig ELLIPSE ---- phi bar circle
%
\linethickness=0.500pt
\setplotsymbol ({\thinlinefont .})
\ellipticalarc axes ratio  0.307:0.307  360 degrees 
	from 9.291 16.669 center at 8.984 16.669
%
% Fig POLYLINE object --- phi bar cross
%
\linethickness=0.500pt
\setplotsymbol ({\thinlinefont .})
\plot 9.143 16.828 8.825 16.510 /
%
% Fig POLYLINE object ---  phi bar cross
%
\linethickness=0.500pt
\setplotsymbol ({\thinlinefont .})
\plot 8.825 16.828 9.143 16.510 /
%
% Fig POLYLINE object ---- phi B line
%
\linethickness=0.500pt
\setplotsymbol ({\makebox(0,0)[l]{\tencirc\symbol{'166}}})
\putrule from 9.302 16.669 to 14.064 16.669
%
% Fig POLYLINE object ---- BB line
%
\linethickness=0.500pt
\setplotsymbol ({\makebox(0,0)[l]{\tencirc\symbol{'166}}})
\putrule from 18.333 16.669 to 23.471 16.669
%
% Fig POLYLINE object ---- del_B cross part
%
\linethickness=1.500pt
\setplotsymbol ({\thinlinefont .})
\plot  4.628 16.510  5.104 16.986 /
%
% Fig POLYLINE object ---- del_B cross part
%
\linethickness=1.500pt
\setplotsymbol ({\thinlinefont .})
\plot  4.628 16.986  5.104 16.510 /
%
% Fig TEXT object
%
\put{$\partial_B$} [lB] at  4.628 14.287
%
% Fig TEXT object
%
\put{$<BB>$} [lB] at 19.400 14.287
%
% Fig TEXT object
%
\put{$<B{\bar\phi}>$} [lB] at 10.204 14.446
\put{${\bar\phi}$} [lB] at 8.530 18.098
\put{$B$} [lB] at 13.828 17.939
\put{$B$} [lB] at 22.976 17.939
\put{$B$} [lB] at 17.579 17.939

\linethickness=0pt
\putrectangle corners at  2.826 26.382 and 36.371 13.161
\endpicture}

\begin{center}
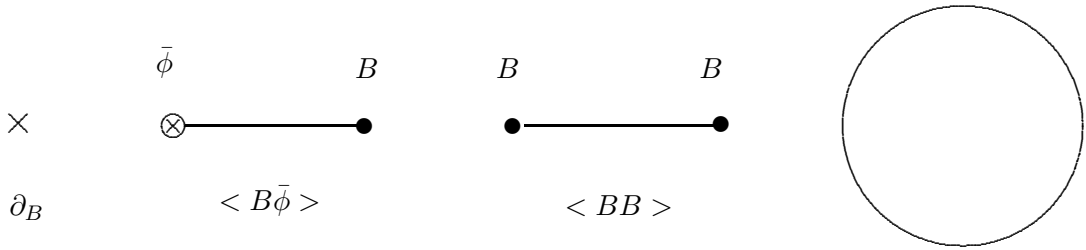
{\bf Figure 1:} The parts for sewing procedure.
\end{center}
\end{minipage}

\vspace{10mm}

\begin{minipage}[t]{13cm}
\font\thinlinefont=cmr5
\begingroup\makeatletter\ifx\SetFigFont\undefined
% extract first six characters in \fmtname
\def\x#1#2#3#4#5#6#7\relax{\def\x{#1#2#3#4#5#6}}%
\expandafter\x\fmtname xxxxxx\relax \def\y{splain}%
\ifx\x\y   % LaTeX or SliTeX?
\gdef\SetFigFont#1#2#3{%
  \ifnum #1<17\tiny\else \ifnum #1<20\small\else
  \ifnum #1<24\normalsize\else \ifnum #1<29\large\else
  \ifnum #1<34\Large\else \ifnum #1<41\LARGE\else
     \huge\fi\fi\fi\fi\fi\fi
  \csname #3\endcsname}%
\else
\gdef\SetFigFont#1#2#3{\begingroup
  \count@#1\relax \ifnum 25<\count@\count@25\fi
  \def\x{\endgroup\@setsize\SetFigFont{#2pt}}%
  \expandafter\x
    \csname \romannumeral\the\count@ pt\expandafter\endcsname
    \csname @\romannumeral\the\count@ pt\endcsname
  \csname #3\endcsname}%
\fi
\fi\endgroup
\mbox{\beginpicture
\setcoordinatesystem units <0.65000cm,0.65000cm>
\unitlength=0.65000cm
\linethickness=1pt
\setplotsymbol ({\makebox(0,0)[l]{\tencirc\symbol{'160}}})
\setshadesymbol ({\thinlinefont .})
\setlinear
%
% Fig CIRCULAR ARC object ----- dash arc in (b)
%
\linethickness=0.500pt
\setshadesymbol ({\thinlinefont .})
%\setdashes < 0.0677cm>
\setdashes < 0.0877cm>
\circulararc 86.117 degrees from 16.034 20.955 center at 18.965 24.262
%
% Fig CIRCULAR ARC object ----- upper dash arc in (b)
%
\linethickness=0.500pt
%\circulararc 93.216 degrees from 19.050 21.749 center at 16.263 18.814
\circulararc 98.858 degrees from 19.050 21.590 center at 16.181 18.996

%
% Fig ELLIPSE ------ big circle in (a)
%
\linethickness=0.500pt
\setplotsymbol ({\thinlinefont .})
\setsolid
\ellipticalarc axes ratio  3.408:3.408  360 degrees 
	from  8.647 21.590 center at  5.239 21.590
%
% Fig ELLIPSE ----- dot in (a)
%
\linethickness=0.500pt
\setplotsymbol ({\makebox(0,0)[l]{\tencirc\symbol{'175}}})
\put{\makebox(0,0)[l]{\circle*{ 0.318}}} at  5.241 24.289
%
% Fig ELLIPSE ----- dot in (a)
%
\linethickness=0.500pt
\setplotsymbol ({\makebox(0,0)[l]{\tencirc\symbol{'175}}})
\put{\makebox(0,0)[l]{\circle*{ 0.318}}} at  5.239 19.050
%
% Fig ELLIPSE
%
\linethickness=0.500pt
\setplotsymbol ({\makebox(0,0)[l]{\tencirc\symbol{'175}}})
\put{\makebox(0,0)[l]{\circle*{ 0.318}}} at 13.176 21.273
%
% Fig ELLIPSE
%
\linethickness=0.500pt
\setplotsymbol ({\makebox(0,0)[l]{\tencirc\symbol{'175}}})
\put{\makebox(0,0)[l]{\circle*{ 0.318}}} at 22.113 21.226
%
% Fig POLYLINE object  ---- bottom cross in (a)
%
\linethickness=0.500pt
\setplotsymbol ({\thinlinefont .})
\plot  5.556 18.415  5.080 17.939 /
%
% Fig POLYLINE object ---- bottom cross in (a)
%
\linethickness=0.500pt
\setplotsymbol ({\thinlinefont .})
\plot  5.080 18.415  5.556 17.939 /
%
% Fig POLYLINE object ----- upper cross in (a)
%
\linethickness=0.500pt
\setplotsymbol ({\thinlinefont .})
\plot  5.397 25.241  4.921 24.765 /
%
% Fig POLYLINE object  ---- upper cross in (a)
%
\linethickness=0.500pt
\setplotsymbol ({\thinlinefont .})
\plot  4.921 25.241  5.397 24.765 /
%
% Fig POLYLINE object  ----- vertical line in (a)
%
\linethickness=0.500pt
\setplotsymbol ({\makebox(0,0)[l]{\tencirc\symbol{'175}}})
\putrule from  5.239 24.130 to  5.239 19.209
%
% Fig POLYLINE object ------- left cross in (b)
%
\linethickness=0.500pt
\setplotsymbol ({\thinlinefont .})
\plot 15.744 21.048 16.192 21.497 /
%
% Fig POLYLINE object  ------ left cross in (b)
%
\linethickness=0.500pt
\setplotsymbol ({\thinlinefont .})
\plot 16.192 21.048 15.744 21.497 /
%
% Fig POLYLINE object ----- right cross in (b)
%
\linethickness=0.500pt
\setplotsymbol ({\thinlinefont .})
\plot 18.919 21.048 19.367 21.497 /
%
% Fig POLYLINE object ----- right cross in (b)
%
\linethickness=0.500pt
\setplotsymbol ({\thinlinefont .})
\plot 19.340 21.048 18.891 21.497 /
%
% Fig POLYLINE object  ------- horizontal line in (b)
%
\linethickness=0.500pt
\setplotsymbol ({\makebox(0,0)[l]{\tencirc\symbol{'175}}})
\putrule from 13.335 21.273 to 21.907 21.273
%
% Fig TEXT object
%
\put{\SetFigFont{12}{14.4}{rm}(a)} [lB] at  5.000 16.351
%
% Fig TEXT object
%
\put{\SetFigFont{12}{14.4}{rm}(b)} [lB] at 17.462 16.510
\linethickness=0pt
\putrectangle corners at  1.814 25.267 and 22.289 16.275
\endpicture}

\begin{center}
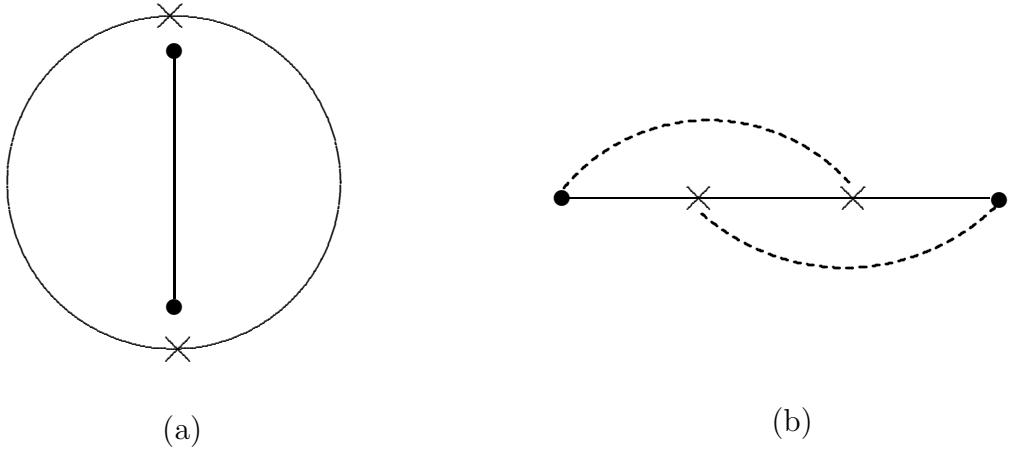
{\bf Figure 2:}  The example of sewing diagrams for $\Delta_{12}^3$.
\end{center}
\end{minipage}
\vspace{15mm}

Now the coefficient $w_i$ can easily be obtained by following 
the above rules, and we remind ourselves of the important fact 
that $w_i$ contains the information on the number of topologies 
of the class in question. Roughly speaking, $w_i$ should be the 
number of topologically independent diagrams of a class $i$, 
multiplied by its symmetry factor. However this is over-counting 
by $N!$ because external legs are not fixed in \eq{zn} yet. 
For example, in the case of the 4-point one-loop diagram, 
$w=1/8$, $S=1$, and $n^T=3$ corresponding to $s$-, $t$-, 
$u$-channels. Therefore the correct relation is  
\beq
w_i = {S_i n^T_i \over N!} \ .   \label{wigeneral}
\eeq

As can easily be seen from the above rules, 
the $w_i$'s for one-loop $N$-point diagrams are universal, 
namely they lead to 
\beq
w_N^{1-loop} = {1\over2N} \ ,  \label{wnoneloop}
\eeq
and the number of topologies can be read from \eq{wigeneral}. 
The two loop cases are more complicated, and we discuss them case 
by case in the following sections. 

%%%%%%%%%%%%%%%%%%  Sect.4 Covering Multiplicity %%%%%%%%%%%%%%%%%%%% 
\section{The covering multiplicity for $N_3=0$ case}
\setcounter{equation}{0}
%%%%%%%%%%%%%%%%%%%%%%%%%%%%%%%%%%%%%%%%%%%%%%%%%%%%%%%%%%%%%%%%%%%%
\indent

In this section, we restrict ourselves to the loop-type 
parametrization with {\it no leg insertions in the middle line} $T_3$. 
The purpose of this section is to give a prescription how to determine 
the covering multiplicity which is the number of times that a world-line 
master integration $\int_M\prod d\tau$ covers all of 
Feynman diagrams $d_n$ within a fixed class $i$. 

Let us begin with the one-loop cases for transparency of discussions. 
Obviously, a symbol diagram $F_i$ for $N$-point diagrams is 
unique for each $N$, and the diagram class may rather be labeled by 
$N$, instead of the original definition of $i$. In the master formula 
\eq{oneloop}, the $\tau$-integration $\int_M$ can be divided into 
$(N-1)!$ integration regions $D_k$, $k=1,\dots,(N-1)!$, 
\beq
\Gamma_M^{1-loop} = \sum_{k=1}^{(N-1)!}\Gamma_{D_k}^{1-loop}\ .
\eeq
Remember the Feynman integral for a certain graph $d_{n'}$ can be 
organized in the following form for some values $k'$ of $k$ 
\beq
    \Gamma[d_{n'}] = S\Gamma_{D_{k'}}^{1-loop} \ . \label{kankei}
\eeq 
(The number of such $k'$-values is equal to the covering multiplicity.) 
Since we must pick up relevant $D_k$'s only one time for each 
topologically different Feynman diagram $d_n$, $n=1,\cdots,n^T$, 
the covering multiplicity in this case is simply given by the ratio 
between the number of the domains and the number $n^T$
\beq
C_N = {(N-1)! \over n^T}\ .  \label{CNdef}
\eeq
Taking account of this covering multiplicity, the naive sum over 
$k=1,\dots,(N-1)!$ can be reduced to the sum only over a 
topologically independent $k$'s subset identical to ${\cal T}_N$ 
\beq
\Gamma_M^{1-loop} = C_N \sum_{k\in{\cal T}_N}\Gamma_{D_k}^{1-loop}
= C_N S_N^{-1}\sum_{n'\in{\cal T}_N}\Gamma[d_{n'}]
= C_N S_N^{-1} {\bar\Gamma}_{F_N} \ ,         \label{GM1loop}
\eeq
where we have used the relation \eq{kankei} at the second equality. 
{}~{}~From \eq{wigeneral} and \eq{wnoneloop}, we have 
\beq
n^T = {(N-1)!\over 2S_N} \ ,
\eeq
and inserting this into \eq{CNdef}
\beq
C_N S_N^{-1}= 2 \ ,
\eeq
we conclude
\beq
\Gamma_N=
{\bar\Gamma}_{F_N} = {1\over2}\Gamma_M^{1-loop}\ . \label{GF1loop}
\eeq
Remember that this factor actually coincides with the one we ignored in 
\eq{oneloop}. 

In the two-loop cases, we have to remember that another 
$\tau$-integration, $\int d\tau_\beta$, exists on the fundamental 
loop, when dividing integration regions. Thus the total number of 
integration domains is $N_M\equiv (N+1)!$, and we have
\beq
\Gamma_M^{(N,0)}=\sum_{k=1}^{N_M}\Gamma_{D_k}^{(N,0)}\ , 
\qquad\mbox{where}\quad N=N_1+N_2 \ . \label{GMN0}
\eeq
Depending on the position of $\tau_\beta$, several different diagram 
classes may appear, and we must count the covering multiplicities 
$C_i$ for the respective diagram classes. To this end, we have 
to classify the $N_M$ regions into $N_{D_i}$ regions for each 
diagram class $i$. 

Suppose $\tau_\beta$ is in such a position that $N_1$ legs are 
positioned on the left hand side (positive $\tau$-direction) of 
$\tau_\beta$, and $N_2$ legs are on the right hand side. 
Then the class label $i$ can be chosen as a set of two integers 
like $F(N_1,N_2)$, $N_1\geq N_2$, and the number of diagram 
classes is given by 
\beq
N_c = [{N\over2}] +1 \ .
\eeq
The number $N_{D_i}$ of integration regions for a class $i=(N_1,N_2)$ 
is equal to  
\beq
N_{D_i} = N! N_p^{(N_1,N_2)} \ ,    \label{NDi}
\eeq
where $N_p^{(N_1,N_2)}$ is the number to deliver the two objects 
$N_1$ and $N_2$ in the two 'post boxes' $(*|*)$; i.e.
\beq
N_p^{(N_1,N_2)}= \left\{\begin{array}{ll}
2   &\quad\mbox{for}\quad   N_1\not= N_2   \\
1   &\quad\mbox{for}\quad   N_1=N_2 \ . \\
\end{array}\right.
\eeq
Eq.\eq{NDi} may also be understood as the number of possible 
constructions of the symbol diagram $F(N_1,N_2)$, namely 
$N_{D_i}=P_{N_1}P_{N_2}N_p^{(N_1,N_2)}$ where 
\beq
P_{N_1}={}_{N}C_{N_1} \times N_1!\ , \qquad P_{N_2} = {}_{N-N_1}C_{N_2}
\times N_2!\ , \qquad {}_NC_r\equiv {N!\over r!(N-r)!}
\eeq
$P_{N_1}$ ($P_{N_2}$) being the number of possibilities to have 
$N_1$ ($N_2$) $\tau$-parameters on the left (right) side of 
$\tau_\beta$ (with over-counting leg orderings). For completeness of 
arguments, we remark that the total sum of $N_{D_i}$ is equal to $N_M$
\beq
\sum_{i=1}^{N_c}N_{D_i}=N!\sum_{i=1}^{N_c}N_p^{(N_1,N_2)}=(N+1)!\ ,
\eeq
where we have used the fact that the number of the classes with 
$N_1>N_2$ is given by $[{N\over2}]$/$[{N+1\over2}]$ and 
the number of the class with $N_1=N_2$ is 1/0 for $N$ even/odd 
respectively. After all, the covering multiplicity $C_i$ for the set 
${\cal T}_i$ with $n^T_i$ diagrams is given by the ratio 
between $N_{D_i}$ and $n^T_i$, thus 
\beq
C_i = {N! N_p^{(N_1,N_2)} \over n^T_i } \ . \label{Cidef}
\eeq

Now, taking account of the multiplicity $C_i$, we rewrite \eq{GMN0} in 
the same way as \eq{GM1loop} for each class
\beq
\Gamma_M^{(N,0)}=
\sum_{i=1}^{N_c}C_i\sum_{k\in{\cal T}_i}\Gamma_{D_k}^{(N,0)} =
\sum_{i=1}^{N_c}C_iS_i^{-1}\sum_{n\in{\cal T}_i}\Gamma[d_n] = 
\sum_{i=1}^{N_c}C_iS_i^{-1}{\bar\Gamma}_{F_i}\ ,
\label{GMdecompose}
\eeq
and an important question here is whether the values of $C_iS_i^{-1}$, 
$i=1,\dots, N_c$, are all the same or not. If they are all the same, the 
unique value gives nothing but a normalization constant for 
$\Gamma_M^{(N,0)}$ just like seen in \eq{GF1loop}. On the other hand, 
if some of $C_iS_i^{-1}$ take a different value, we have to add a 
correction to make the summation weights over $i$ equal. 
Eliminating $n^T_i$ from 
the $C_i$ \eq{Cidef} by using \eq{wigeneral}, we derive
\beq
C_iS_i^{-1}={N_p^{(N_1,N_2)} \over w_i} \label{CSformal}\ ,
\eeq
and we have verified (so far $N\leq5$) that its explicit values are 
\beq
C_iS_i^{-1}= \left\{\begin{array}{ll}
8  &\quad\mbox{for}\quad   (N_1,N_2)=(N,0)   \\
4  &\quad\mbox{otherwise}\quad   \ . \\
\end{array}\right. \label{CSdata}
\eeq
Here, a few remarks are in order: (i) 
On the ground that graphical symmetries are getting less as $N$ is 
increasing, we believe that the data obtained above should be 
valid for larger $N$-values as well. (ii) 
The main reason why the $CS^{-1}$ value for the $F(N,0)$ 
class is twice that for the rest is the fact that the symmetry 
factor of $F(N,0)$ is nothing but $1/2$. (The $N=2$ case is the only 
exception, where both symmetry factors are $1/2$, but $C$'s are 
different). (iii) If the fundamental loop has an orientation like a 
fermion loop in QED, the symmetry factor for $F(N,0)$ actually turns 
to be 1, and hence $CS^{-1}=C=2$ for {\it all} classes, where we have 
divided $C$ by 2 because of distinguishing different directions of 
the orientated loop. In other words, $n^T_i$ of the denominator in 
\eq{Cidef} is twice as large as the one for the $\phi^3$ theory. 
(The $N=2$ case is again exceptional.)  

Now applying the results \eq{CSdata} to the formula \eq{GMdecompose}, 
we find 
\beq
\Gamma_M^{(N,0)} = 4\sum_{i=1}^{N_c}{\bar\Gamma}_{F_i}
+ 4 {\bar\Gamma}_{F(N,0)} \ .
\eeq
The quantity appearing in the first term of the above equation 
is nothing but the $N$-point function in question
\beq
\ogamma_N = \sum_{i=1}^{N_c}{\bar\Gamma}_{F_i} \ ,
\eeq
and the second term is given by the following master formula 
relation, which will be proven in the next section
\beq
{\bar\Gamma}_{F(N,0)}={\bar\Gamma}_{F(N,0,0)}=
{1\over4}\Gamma_M^{(N,0,0)} \ .    \label{prop1}
\eeq
Therefore we arrive at 
\beq
\ogamma_N ={1\over4}(\Gamma_M^{(N,0)}-\Gamma_M^{(N,0,0)})\ . 
\label{main1}
\eeq

In view of the remark (iii) below \eq{CSdata}, 
the $N$-point amplitude of (two-loop) QED photon scatterings is 
simply given by  
\beq
\Gamma_N = {1\over2}\Gamma_M^{QED}\ , \label{QEDampli}
\eeq
with
\beqa
\Gamma_M^{QED} &=& -4\cdot {e^{N+2}\over(4\pi)^D} 
\int_0^\infty{dT\over T}\int_0^\infty d{\bar T} 
\int\prod_{n=1}^N d{\hat\tau}_n d{\hat\tau}_\alpha d{\hat\tau}_\beta
[T{\bar T}+T{\hat G}({\hat\tau}_\alpha,{\hat\tau}_\beta)]^{-D/2}\nn\\
&\times&<D{\hat X}({\hat\tau}_\alpha)\cdot
D{\hat X}({\hat\tau}_\beta)\prod_{n=1}^N
D{\hat X}({\hat\tau}_n)\cdot\epsilon_n\exp[ip_n\cdot{\hat X}({\hat\tau}_n)]>
\label{QEDmaster}
\eeqa
where $D=\partial_\theta-\theta\partial_\tau$, and 
${\hat X}_\mu({\hat\tau}_n)$, $n=1,\cdots,N, \alpha,\beta$ have to 
be contracted with the Green function ${\hat G}^{(1)}_{11}({\hat\tau}_j, 
{\hat\tau}_k)$ obtained from the $G^{(1)}_{11}(\tau_j,\tau_k)$ in 
\eq{masterL} by substitution of $G_B$ with a super-Green 
function~\cite{SSqed} 
${\hat G}({\hat\tau}_1,{\hat\tau}_2)=G_B(\tau_1,\tau_2)+
\theta_1\theta_2 G_F(\tau_1,\tau_2)$, where $G_F(\tau_1,\tau_2)=
\mbox{sign}(\tau_1-\tau_2)$. The factor $-4$ appearing in 
$\Gamma_M^{QED}$ corresponds to the fermion degrees of freedom. 
For scalar QED, just reduce this number to 
$2$, also reduce super variables to bosonic ones: 
$D{\hat X}\ra {\dot x}$, ${\hat\tau}\ra\tau$ \cite{SSqed,sumino}. 

As shown in Appendix B, one may fix one of the $\tau$-integrations 
by using the translational invariance to perform the $\tau$-integration trivially. If one wants to fix $\tau_\alpha=0$, then we replace 
\beq
\int_0^T d{\hat\tau}_{\alpha} \quad\ra\quad T \int d\theta_\alpha \ .
\eeq

%%%%%%%%%%%% Sect.5 Combinatorics in the symmetric formula %%%%%%%%%%%%% 
\section{Combinatorics in the symmetric master formula}
\setcounter{equation}{0}
%%%%%%%%%%%%%%%%%%%%%%%%%%%%%%%%%%%%%%%%%%%%%%%%%%%%%%%%%%%%%%%%%%%%
\indent

In the symmetric parametrization, the following proper $N$-point 
function can be derived from $Z[{\bar\phi}]$ (see Appendix A)  
\beq
\Gamma_N = {1\over 2\cdot 3!}
\sum_{N_1,N_2,N_3=0}^N
\sum_{\sigma(N_1,N_2,N_3)}\Gamma_M^{(N_1,N_2,N_3)}\ ,
\qquad (N=N_1+N_2+N_3) \label{newGamma}
\eeq
where $\sigma$ stands for possible ways of inserting $N_a$ particles 
(momenta) in the internal three lines ignoring leg orderings on each 
line. To perform these summations efficiently, we restrict 
$N_1\geq N_2\geq N_3$ and take account of the multiplicity 
$N_p^{(N_1,N_2,N_3)}$;
\beq
\sum_{N_1,N_2,N_3}^N \ra \sum_{N_1\geq N_2\geq N_3}^N 
N_p^{(N_1,N_2,N_3)}  \label{sumreduce}
\eeq
where 
\beq
N_p^{(N_1,N_2,N_3)} = \left\{\begin{array}{ll}
1  &\quad\mbox{if all of $N_a$ are the same}\quad   \\
3  &\quad\mbox{if two of $N_a$ are the same}\quad   \\
6  &\quad\mbox{if none of $N_a$ is the same}\ . \\
\end{array}\right. \label{Npnml}
\eeq
However this is not the end of the story. There comes another kind of 
multiplicity from the summation in $\sigma$. The number of $\sigma$'s 
is given by 
\beq
N_\sigma = {}_NC_{N_1}\cdot {}_{N-N_1}C_{N_2}\cdot 
{}_{N-N_1-N_2}C_{N_3} = {N!\over N_1!N_2!N_3!}\ ,
\eeq
and we show some examples of $\sigma(N_1,N_2,N_3)$ with 
$N_1\geq N_2\geq N_3$,  
\beqa
&&\sigma(1,0,0) =(p_1|0|0), \quad \sigma(2,0,0) =(p_1,p_2|0|0),
\quad \sigma(3,0,0) =(p_1,p_2,p_3|0|0)     \label{exam1}  \\
&&\sigma(1,1,0) =(p_1|p_2|0), \ (p_2|p_1|0)\\
&&\sigma(2,1,0) =(p_1,p_2|p_3|0), \ (p_1,p_3|p_2|0), \ (p_2,p_3|p_1|0) \\
&&\sigma(1,1,1) =(p_1|p_2|p_3), \ (p_1|p_3|p_2), \ (p_2|p_1|p_3),\ 
(p_2|p_3|p_1),\ \dots\quad\mbox{etc.}, \label{exam2}
\eeqa
where a momentum $p_k$ delivered in the $a$-th place of $(*|*|*)$ is 
$p_k^{(a)}$ appearing in the master formula \eq{masterS}. 
Because of \eq{sumreduce}, we find out the multiplicities 
2 for $\sigma(1,1,0)$, 6 for $\sigma(1,1,1)$ and 1 for the remainders in 
these examples \eq{exam1}-\eq{exam2}. A more general argument for these 
multiplicities $C_{N_1N_2N_3}$ leads us to the following relations 
\beqa
&&N_p^{(N_1N_2N_3)}C_{N_1N_2N_3} = 6 \qquad\mbox{for}\quad N_3\not=0, \\
&&N_p^{(N_1N_2)}C_{N_1N_2N_3} =2 \qquad\mbox{for}\quad N_3=0.
\eeqa
Noticing the relation 
\beq
{ N_p^{(N_1N_2)}\over N_p^{(N_1N_2N_3)}} = 
{C_iS_i^{-1}\over 2\cdot3!}, \qquad\mbox{for}\quad N_3=0, 
\quad i=(N_1,N_2) \label{NNrelation}
\eeq
we have 
\beq
N_p^{(N_1N_2N_3)}C_{N_1N_2N_3}= 6(1-\delta_{N_3,0})
+{24\over C_iS_i^{-1}}\delta_{N_3,0} \ . \label{NPC}
\eeq
Then \eq{newGamma} turns out to be 
\beq
\Gamma_N = {1\over2\cdot3!} \sum_{N_1\geq N_2\geq N_3}^N
N_p^{(N_1N_2N_3)}C_{N_1N_2N_3}{\bar\Gamma}_M^{(N_1,N_2,N_3)}\ , 
\label{main2}
\eeq
where
\beq
{\bar\Gamma}_M^{(N_1,N_2,N_3)}=\sum_\sigma{}\hskip-0.5mm'\,
\Gamma_M^{(N_1,N_2,N_3)}\ , \label{permutateGamma}
\eeq
and $\sum'$ means that redundant elements of $\sigma$ have been 
subtracted according to the multiplicities $C_{N_1N_2N_3}$. First 
few examples are 
\beqa
&&{\bar\Gamma}_M^{(1,1,0)} =\Gamma_M^{(1,1,0)}(p_1|p_2|0)\ , \\
&&{\bar\Gamma}_M^{(1,1,1)} =\Gamma_M^{(1,1,1)}(p_1|p_2|p_3)\ , \\
&&{\bar\Gamma}_M^{(2,1,1)}= \sum_{i<j}^4\Gamma_M^{(2,1,1)}
(p_1,..., {\check p}_i,..., {\check p}_j,... p_4|p_i|p_j) \ ,
\eeqa
where ${\check p}_j$ means an exclusion of $p_j$. 

Now, let us prove \eq{prop1}. Recalling that the $\Gamma_N$ should 
be a sum over all classes labeled by $(N_1,N_2,N_3)$, 
$N_1\geq N_2\geq N_3$, 
\beq
\Gamma_N= \sum_i{\bar\Gamma}_{F_i} = \sum_{N_1\geq N_2\geq N_3}
{\bar \Gamma}_{F(N_1,N_2,N_3)} \ ,
\eeq
and comparing this with \eq{main2}, we find  
\beq
{\bar\Gamma}_{F(N_1,N_2,N_3)} = {1\over2\cdot3!}\cdot
N_p^{(N_1N_2N_3)}C_{N_1N_2N_3}\cdot{\bar\Gamma}_M^{(N_1,N_2,N_3)} \ .
\label{mapFtoSM}
\eeq
In the case of $(N,0,0)$, noticing 
${\bar\Gamma}_M^{(N,0,0)}=\Gamma_M^{(N,0,0)}$, we obtain \eq{prop1} 
as 
\beq
{\bar \Gamma}_{F(N,0,0)} = {1\over2\cdot3!}\cdot
{24\over8}\cdot\Gamma_M^{(N,0,0)} = {1\over4}\Gamma_M^{(N,0,0)}\ .
\eeq

Here we also derive several interesting formulae. 
Since we already know the result \eq{main1} which contains 
every class such as $(N_1,N_2,0)$, we can {\it partly} perform 
the summation over $N_1$ and $N_2$ in \eq{main2}. 
Using \eq{CSdata} and \eq{NPC}, we have 
\beq
\Gamma_N = \ogamma_N 
+{1\over2}\sum_{N_1\geq N_2\geq N_3>0}^{N=N_1+N_2+N_3} 
{\bar\Gamma}_M^{(N_1,N_2,N_3)} \ , \label{partsum1}
\eeq
where we see that $\ogamma_N$ is given by
\beq
\ogamma_N = \sum_{N_1\geq N_2}^{N=N_1+N_2}{2\over CS^{-1}} 
{\bar\Gamma}_M^{(N_1,N_2,0)} \ . \label{partsum2}
\eeq
Comparing this $\ogamma_N$ with \eq{main1}, we can derive a 
decomposition formula of $\Gamma_M^{(N,0)}$ in terms of 
${\bar\Gamma}^{(N_1,N_2,0)}$ 
\beq
{1\over2}\Gamma_M^{(N,0)} =\sum_{N_1\geq N_2}
{\bar\Gamma}^{(N_1,N_2,0)} \ . \label{GMtransform}
\eeq

As a closing remark, we suggest a simple relation between 
$w_i$ and $N_p^{(N_1N_2N_3)}$. Eqs. \eq{CSformal} and 
\eq{NNrelation} are defined only for $N_3=0$; however 
eliminating $N_p^{(N_1,N_2)}$ from \eq{CSformal} with 
$N_3$ being kept finite formally, we find 
\beq
w_i = {N_p^{(N_1N_2N_3)}\over2\cdot3!}\ . \label{simplewi}
\eeq
We verified, up to $N=6$, that this result is correct also for 
non zero $N_3$ values. 

%%%%%%%%%%%%%%%%%%%% Sect. 6 %%%%%%%%%%%%%%%%%%%%%%%%%%%%%%%%%%%%%%%%
\section{Non zero $N_3$ case in the loop type formulae}
\setcounter{equation}{0}
\indent
%%%%%%%%%%%%%%%%%%%%%%%%%%%%%%%%%%%%%%%%%%%%%%%%%%%%%%%%%%%%%%%%%%%%%

We, in principle, expect a generic method to evaluate combinatorics 
like discussed in sect.4 in the non zero $N_3$ case of the loop 
type master formula. Here we confine ourselves to a few 
examples as a first step. 

The simplest case is $F(1,1,1)$, which can be covered by the 
master formula $\Gamma_M^{(2,1)}$. This case does not give rise to 
any complication from sect. 4, because of only one insertion 
to the {\it middle line} propagator. Since $\Gamma_M^{(2,1)}$ 
covers the two classes $F(2,1,0)$ having $C=4$, $S=1$ and 
$F(1,1,1)$ having $C=2$, $S=1/2$, we get
\beq
\Gamma_M^{(2,1)} = 4{\bar\Gamma}_{F(2,1,0)} + 
4\Gamma_{F(1,1,1)}\ , 
\eeq
where we have used ${\bar\Gamma}_{F(1,1,1)}=\Gamma_{F(1,1,1)}$. 
One may obtain the connection of $\Gamma_M^{(2,1)}$ to the symmetric 
master formula by applying \eq{mapFtoSM} to the above equation
\beq
{1\over2}\Gamma_M^{(2,1)} = {\bar\Gamma}_M^{(2,1,0)} +
 \Gamma_M^{(1,1,1)}\ .  \label{combi1}
\eeq
{}~From \eq{GMtransform} we also have 
\beq
{1\over2}\Gamma_M^{(3,0)} = \Gamma_M^{(3,0,0)}+
{\bar\Gamma}_M^{(2,1,0)} \ . \label{combi2}
\eeq
These two formulas give a transformation from \eq{main2}, the 
sum of symmetric master formulae, to a loop type formula for 
$\Gamma_3$ Writing down \eq{main2} for $N=3$
\beq
\Gamma_3 = {1\over4}\Gamma_M^{(3,0,0)}+{1\over2}{\bar\Gamma}_M^{(2,1,0)}
+{1\over2}\Gamma_M^{(1,1,1)}\ , \label{Gamma3}
\eeq
and taking a linear combination of \eq{combi1} and \eq{combi2}, 
we can rewrite \eq{Gamma3} in the following form
\beq
\Gamma_3 = {1\over8}\Gamma_M^{(3,0)}+ {1\over8}\Gamma_M^{(2,1)}
+{1\over4}\Gamma_M^{(1,1,1)} \ . \label{loopgamma3}
\eeq
This formula is an interesting form; notice that it does not contain 
any summation in momentum permutations. (This is also true 
for \eq{partsum1}.)

Now, the second simplest example is $F(2,1,1)$, which contains 6 
topologically different Feynman diagrams in the class. This class 
is expected to be covered by $\Gamma_M^{(3,1)}$, and $N_3$ 
is again equal to one, thus the situation may be similar to the 
first case. The number of sub-regions is determined by orderings 
of $\tau_\beta$ and $\tau_n$ on the {\it fundamental loop}, i.e. 
$(N_1+N_2+1)!=4!$. However these sub-regions can not cover 
all 6 diagrams of ${\cal T}(2,1,1)$. In fact, only 3 diagrams among 
them are covered with the covering multiplicity 4, 
and the remaining 12 regions cover 3 diagrams belonging 
to ${\cal T}(3,1,0)$ with multiplicity 4. In order to cover all 
necessary diagrams of ${\cal T}(2,1,1)$, we just permutate 
the momentum $p_j$ inserted in the {\it middle line} and gather 
them all for $j=1,2,3,4$. Thus we obtain
\beqa
{\bar\Gamma}_M^{(3,1)}\equiv
\sum_{j=1}^4 \Gamma_M^{(3,1)}(p_1,..,{\check p}_j,..,p_4|p_j) 
&=& 8{\bar\Gamma}_{F(2,1,1)} + 4{\bar\Gamma}_{F(3,1,0)} \label{G31def}\\
&=& 2{\bar\Gamma}_M^{(3,1,0)}+4{\bar\Gamma}_M^{(2,1,1)}\ ,
\label{combi3}
\eeqa
where we have used \eq{mapFtoSM} at the second equality. 
{}~From \eq{GMtransform} and \eq{main2}, we have 
\beq
{1\over2}\Gamma_M^{(4,0)}=\Gamma_M^{(4,0,0)}+
{\bar\Gamma}_M^{(3,1,0)}+{\bar\Gamma}_M^{(2,2,0)} \label{combi4}
\eeq
and 
\beq
\Gamma_4={1\over4}\Gamma_M^{(4,0,0)}+{1\over2}{\bar\Gamma}_M^{(3,1,0)}
+{1\over2}{\bar\Gamma}_M^{(2,2,0)}+{1\over2}{\bar\Gamma}_M^{(2,1,1)}\ .
\eeq
We can rewrite this $\Gamma_4$ into the following form by taking 
a linear combination of \eq{combi3} and \eq{combi4}
\beq
\Gamma_4={1\over8}\Gamma_M^{(4,0)}+{1\over8}{\bar\Gamma}_M^{(3,1)}
+{1\over4}{\bar\Gamma}_M^{(2,2,0)} \ .\label{loopgamma4}
\eeq
This result contains summations in external momenta's permutations, 
because of \eq{G31def}, the definition of ${\bar\Gamma}_M^{(3,1)}$, and 
\beq
{\bar\Gamma}_M^{(2,2,0)} =\Gamma_M^{(2,2,0)}(p_1,p_2|p_3,p_4|0)+
\Gamma_M^{(2,2,0)}(p_2,p_4|p_1,p_3|0)+
\Gamma_M^{(2,2,0)}(p_2,p_3|p_1,p_4|0) \ .
\eeq
However the number of summations is much less than in the standard 
Feynman rule calculation, which needs a sum of 36 diagrams. 
The contents of 36 diagrams are 12, 12, 6, 6 for ${\cal T}(4,0,0)$, 
${\cal T}(3,1,0)$, ${\cal T}(2,2,0)$, ${\cal T}(2,1,1)$ respectively. 

In summary we have obtained the simple expressions for $\Gamma_3$ and 
$\Gamma_4$. $\Gamma_3$ consists of only three $\Gamma_M$'s in 
\eq{loopgamma3}, while the Feynman rules need 3 diagrams (permutations) 
for ${\cal T}(3,0,0)$ and ${\cal T}(2,1,0)$ each, and one diagram 
for ${\cal T}(1,1,1)$. For $\Gamma_4$ in \eq{loopgamma4} we only 
need the summation of $\Gamma_M^{(4,0)}$, 4 permutations of 
$\Gamma_M^{(3,1)}$, and 3 permutations of $\Gamma_M^{(2,2,0)}$. 
These simplifications can be seen in \eq{partsum1} as well. 

%%%%%%%%%%%%%%%%%%%% Sect. 7 %%%%%%%%%%%%%%%%%%%%%%%%%%%%%%%%%%%%%%%%
\section{Conclusions}
\setcounter{equation}{0}
\indent
%%%%%%%%%%%%%%%%%%%%%%%%%%%%%%%%%%%%%%%%%%%%%%%%%%%%%%%%%%%%%%%%%%%%%

In this paper we determined the correct normalizations and combinatorics 
of two-loop world-line formulae for proper $N$-point Feynman amplitudes 
in $\phi^3$-theory and QED photon scatterings. For $\phi^3$-theory the 
full result is given by \eq{partsum1}, where the first term $\ogamma_N$ 
contains the sum over all Feynman diagrams for $N_3=0$ and can be 
obtained by only two quantities as shown in \eq{main1}. The second term 
of \eq{partsum1} is also a sum over the remaining Feynman diagrams, 
where basically we have only to calculate one master formula for each 
class, then gather all permutations defined in \eq{permutateGamma}. 

Let us see how simple these results are in the cases of $N=4$ and 5, 
where the numbers of Feynman diagrams are 36 and 240 (These numbers 
can be obtained as a sum of all $n^T_i$ given by \eq{Sfactor}, 
\eq{wigeneral} and \eq{simplewi}). Among those diagrams, 30 and 180 
diagrams are encapsulated in the $\ogamma_N$ respectively. Then 
for $N=4$ we are left with 6 permutations 
of $\Gamma^{(2,1,1)}_M$, and for $N=5$ there are 15 permutations of 
$\Gamma^{(2,2,1)}_M$, 10 permutations of $\Gamma^{(3,1,1)}_M$ 
and no more. One might think that these are still many summations. 

However this problem is unavoidable as far as we adopt the world-line 
parametrizations which do not really parametrize two loop cycles. For 
example, one can see that our loop type formulae \eq{loopgamma3} and  \eq{loopgamma4} were able to absorb symmetric type quantities to some 
extent, but some momentum permutations still remain in \eq{loopgamma4}. 
This is caused by the fact that a vertex on the {\it middle line} can 
not move beyond the joining point along the second loop cycle, while 
a vertex on the fundamental loop (first loop cycle) can move in the 
entire loop. Actually from a string theoretical viewpoint, the middle 
line proper-time variable is defined by the difference between two 
points $\tau^{(\alpha)}$ and $\tau^{(3)}$ along the second loop 
cycle~\cite{RS1}. In order to solve this problem, one should really 
formulate the middle line as a loop, where the new formulation 
might look alike a string theory more than ever.

Fortunately in the case of QED photon scatterings, our loop 
parametrization is sufficient to gather all the Feynman diagrams in 
the single master formula \eq{QEDampli}~\cite{SSqed}. Therefore 
a multi-loop $N$-point formula will also be obtainable in the same way 
as discussed in sect. 4. If the covering multiplicity will be given 
by $C_h$ for the $(h+1)$-loop master integration region 
$M_h=\{ 0\leq\tau_a\leq T|\, a=1,\dots,N,\alpha_i,
\beta_j;\, i=2,\dots,h;\, j=1,\dots,h\}$, the $(h+1)$-loop $N$-point 
Feynman amplitudes are given by 
\beqa
\Gamma^{(h+1)}_N&=& -4C_h^{-1}(4\pi)^{-{D\over2}(h+1)}e^{N+2h}
\int_0^\infty dTT^{-D/2}\prod_{i=1}^h\int_0^\infty d{\bar T}_i\cdot 
\int d\theta_{\alpha_1}   \nn\\
&\times&\int_{M_h}d{\hat\tau}_{\beta_1}\prod_{i=2}^h d{\hat\tau}_{\alpha_i}
d{\hat\tau}_{\beta_i}\prod_{n=1}^Nd\tau_n \cdot(\mbox{det}{\hat A})^{-D/2}\\
&\times&\left.<\prod_{i=1}^hD{\hat X}({\hat\tau}_{\alpha_i})\cdot
D{\hat X}({\hat\tau}_{\beta_i})\prod_{n=1}^N D{\hat X}({\hat\tau}_n)
\cdot\epsilon_n\exp[ip_n\cdot{\hat X}({\hat\tau}_n)]>
\right|_{\tau_{\alpha_1}=0}\nn
\eeqa
where $\mbox{det}{\hat A}$ is the determinant defined by switching 
to super Green functions~\cite{SSqed} in the determinant factor 
$\mbox{det}A$ appearing in the multi-loop $\phi^3$-theory 
formula~\cite{SSphi,RS1}. 

It is valuable to notice in the world-line formulation that the 
generating functional of $N$-point 1PI amplitudes takes a very 
simple and compact form, where the fundamental loop and all 
internal propagators are naturally joined by an auxiliary field 
(either $B$ or $\alpha$) at an {\it arbitrary} loop order --- 
though we only demonstrated the two-loop cases in sect. 3. 
In the two-loop cases, this may rather be trivial from the 
viewpoint of three propagators convolution in a background; 
however the other version with a circle world-line is non-trivial 
in the spirit of multi-loop generalization \cite{SSphi}. 

It will also be interesting to apply our methods to a non-abelian 
gauge theory and $\phi^4$ theory ~\cite{xx}. As discussed in sect.3, 
we should first derive a simple expression for the (two-loop) 
effective action based on world-line expressions for the 
propagator and fundamental loop in a background. Writing the 
background field as a sum of plane waves and expanding it to an 
appropriate order one will obtain the $N$-point amplitudes 
expressed in terms of a master formula with correct combinatorics 
(as advocated in sect. 5 and Appendix A). Thus one would 
implement a compact form for 1PI amplitudes. 
 
%%%%%%%%%%%%%%%%%%%%%%%%%%%%%%%%%%%%%%%%%%%%%%%%%%%%%%%%%%%%%%%%%%%%%%%%
\appendix

\section*{Appendix A. Derivation of (5.1) from $Z[{\bar\phi}]$}
\setcounter{equation}{0}
\setcounter{section}{1}

Showing a derivation of \eq{newGamma} from \eq{generate}, 
we briefly explain the origin of the overall factor and combinatorics in 
\eq{newGamma}. Since we discuss the proper diagram parts, we may omit 
the tree part $Z_0$. The quadratic terms in $\varphi$ in 
\eq{generate} can be read as the one-loop effective action 
\beq
\Gamma^{1-loop} = -{1\over2}\ln{\rm Det} (-\partial^2+m^2+g{\bar\phi})\ ,
\eeq
and the remainder in \eq{generate} can be interpreted as the internal 
$\varphi^3$ vertex. We have only to pick up the second order in this 
vertex in order to make two-loop diagrams. Therefore the desired 
two-loop contributions are given by 
\beq
\Gamma^{2-loop}= {g^2\over2\cdot(3!)^2}  
\int{\cal D}\varphi d^Dx_1d^Dx_2 \varphi^3(x_1) \varphi^3(x_2)
\exp\Bigl[-{1\over2}\int\varphi(\Delta^{-1}+g{\bar\phi})
\varphi d^Dx\Bigr]\ .
\eeq
Applying Wick contractions we rewrite
\beq
\Gamma^{2-loop} = {g^2\over2\cdot3!} \int d^Dx_1d^Dx_2
<x_1|(\Delta^{-1}+g{\bar\phi})^{-1}|x_2>^3\ . \label{wickgamma}
\eeq
We also know the path integral expression for the propagator
\beq
<x_1|(\Delta^{-1}+g{\bar\phi})^{-1}|x_2>
=\int_0^\infty dT \int_{\scriptstyle y(0)=x_2 
        \atop\scriptstyle y(T)=x_1}{\cal D}y(t)\exp\Bigl[ 
-\int_0^Td\tau({1\over4}{\dot y}^2(\tau)+m^2+g{\bar\phi}(y(\tau))
\Bigr]\ ,  \label{pathprop}
\eeq
where the path integral normalization is given by
\beq
 \int_{\scriptstyle y(0)=x_2 
        \atop\scriptstyle y(T)=x_1}{\cal D}y(\tau)\exp\Bigl[ 
-\int_0^Td\tau {1\over4}{\dot y}^2(\tau) \Bigr]
=(4\pi T)^{-D/2}\exp\Bigl[-{(x_1-x_2)^2\over4T}\Bigr] \ .
\eeq
Substituting this into \eq{wickgamma} and expanding the background 
field, we have 
\beqa
\Gamma^{2-loop}&=&
{g^2\over2\cdot3!} \sum_{N_1,N_2,N_3=0}^\infty
{(-g)^{N_1+N_2+N_3}\over N_1!N_2!N_3!}
\int d^Dx_1d^Dx_2 \prod_{a=1}^3\int_0^\infty dT_a e^{-m^2T_a} \label{Gammaplane}\\
&\times&\int_{\scriptstyle y_a(0)=x_2 
        \atop\scriptstyle y_a(T_a)=x_1}{\cal D}y_a(\tau)
\exp\Bigl[-\int_0^{T_a}{1\over4}{\dot y}_a^2(\tau^{(a)}) d\tau^{(a)}\Bigr]
\Bigl[\int_0^{T_a}{\bar\phi}(y_a(\tau^{(a)})d\tau^{(a)}\Bigr]^{N_a}\ .\nn
\eeqa
Now let us consider plane wave expansions of the background scalar 
vertex operators
\beq
    \phi(y) = \sum_{k=1}^N e^{ip_ky} \ .
\eeq
If we naively insert this expansion, various terms will appear. 
As shall be seen after performing the coordinate integrations, 
we implicitly have a delta function for the total momentum conservation 
in \eq{Gammaplane}. {}~From this reason, we ignore the terms which 
include the same momentum twice after the plane wave substitutions. 
Introducing the following notation
\beq
V^{(a)}_k = \int_0^{T_a}d\tau^{(a)}\exp\Bigl[ip_ky_a(\tau^{(a)})\Bigr]\ ,
\eeq
we are thus allowed to perform the replacement
\beq
\Bigl[\int_0^{T_a}{\bar\phi}(y_a(\tau^{(a)})d\tau^{(a)}\Bigr]^{N_a}
\ra N_a!\sum_{i_1<i_2<\cdots<i_{N_a}} 
V^{(a)}_{i_1} \cdots V^{(a)}_{i_{N_a}}\ , \label{replaceV}
\eeq
where every $i_k$, $k=1,\dots,N_a$ runs from 1 to $N$ as far as the 
ordering restriction is satisfied. The number of such terms is 
${}_NC_{N_a}$. Suppose we performed \eq{replaceV} for $a=1$ and 
are on the verge of applying \eq{replaceV} to the next $a=2$, 
then we should note that $i_k$ should run among $N-N_1$ integers 
this time. Thus the number of terms is ${}_{N-N_1}C_{N_2}$. 
These numbering for external momenta is nothing but the distribution 
$\sigma(N_1,N_2,N_3)$ defined at \eq{newGamma}, and 
we realize that \eq{Gammaplane} for fixed $N$ turns out to be 
\beqa
\Gamma^{2-loop}_N&=&
{(-g)^{N+2}\over2\cdot3!} \sum_{N_1,N_2,N_3}^N\sum_\sigma
\int d^Dx_1d^Dx_2 \prod_{a=1}^3\int_0^\infty dT_a e^{-m^2T_a}\nn\\
&\times&\int_{\scriptstyle y_a(0)=x_2 
        \atop\scriptstyle y_a(T_a)=x_1}{\cal D}y_a(\tau)
\exp\Bigl[-\int_0^{T_a}{1\over4}{\dot y}_a^2 d\tau^{(a)}\Bigr]
\prod_{n=1}^{N_a} \int_0^{T_a}d\tau_n^{(a)}e^{ip_n^{(a)}y(\tau_n^{(a)})}\ .
\label{Gammaplanenew}
\eeqa
First performing the $y$ integrations for example putting
\beq
y_a(\tau)=x_1+{\tau\over T_a}(x_2-x_1)+\sum_{m=1}^\infty
y_m {\rm sin}\Bigl({m\pi\tau\over T_a}\Bigr) \ ,
\eeq
and secondly performing $x$ integrations, 
we finally obtain
\beq
\Gamma^{2-loop}_N={1\over2\cdot3!}\sum_{N_1,N_2,N_3=0}^N
\sum_\sigma
(2\pi)^D\delta(\sum_{a=1}^3\sum_{n=1}^{N_a}p_n^{(a)})
\Gamma_M^{(N_1,N_2,N_3)}\ ,
\eeq
and this coincides with \eq{newGamma} up to the 
$(2\pi)^D\delta(\sum p_n)$. 

%%%%%%%%%%%%%%%%%%%%%%%%%%%%%%%%%%%%%%%%%%%%%%%%%%%%%%%%%%%%%%%%%%%%%%%%%%%%%
\section*{Appendix B. Translational invariance along the fundamental loop}
\setcounter{equation}{0}
\setcounter{section}{2}
%%%%%%%%%%%%%%%%%%%%%%%%%%%%%%%%%%%%%%%%%%%%%%%%%%%%%%%%%%%%%%%%%%%%%%%%%%%%%%

In this appendix, we show how to fix one of the super world-line 
${\hat\tau}$-parameters in view of the invariance of the integrand 
in \eq{QEDmaster} under the translation 
\beq
\tau_n \ \rightarrow \ \tau_n + c \quad \mbox{for} \quad
n=1,\ldots,N, \alpha, \beta \ .\label{shift}
\eeq
This translation simply follows from the property
\beq
{\hat G}({\hat\tau}_a,{\hat\tau}_b) = 
{\hat G}({\hat\tau}_a+c,{\hat\tau}_b+c) \ ,
\eeq
since ${\hat G}_{11}$ is made only of ${\hat G}$. 
 
If we introduce a simplified notation, where 
${\hat x}_0,{\hat x}_1,\ldots,{\hat x}_n$ denote the $n+1=N+2$ super 
world-line ${\hat\tau}$-parameters of the fundamental loop, and the 
integrand of the formula \eq{QEDmaster} is called 
${\hat f}({\hat x}_0,{\hat x}_1,\ldots,{\hat x}_n)$,
then the invariance \eq{shift} allows us to rewrite the amplitude
\eq{QEDampli} as
\beqa
{\cal A}_N &=&\int_0^T\mbox{d}{\hat x}_0\int_0^T\mbox{d}{\hat x}_1\ldots
            \int_0^T\mbox{d}{\hat x}_n\ 
{\hat f}({\hat x}_0,{\hat x}_1,\ldots,{\hat x}_n) \\
&=&\int_0^T\mbox{d}{\hat x}_0\int_0^T\mbox{d}{\hat x}_1\ldots
    \int_0^T\mbox{d}{\hat x}_n \ 
{\hat f}({\hat 0},{\hat x}_1-x_0,\ldots,{\hat x}_n-x_0) \nn \\
&=&\int_0^T\mbox{d}{\hat x}_0\int_{-x_0}^{-x_0+T}\mbox{d}{\hat x}_1\ldots
\int_{-x_0}^{-x_0+T} \mbox{d} {\hat x}_n \ 
{\hat f}({\hat 0},{\hat x}_1,\ldots,{\hat x}_n) \ , \nn
\eeqa
where we have denoted the following fact by ${\hat 0}=(0, \theta_0)$ that 
the dependence on $x_0$ has disappeared from the integrand, however the 
dependence on $\theta_0$ still remains in ${\hat f}$. 

If ${\hat f}$ was periodic with period $T$ in each variable, we could 
just replace the integration region
\beq
\int_{-x_0}^{-x_0+T}\mbox{d}{\hat x}_i \ra \int_0^T\mbox{d}{\hat x}_i
\label{replace}
\eeq
for each $i=1,\ldots,n$. However the rigorous situation is not 
straightforward because of the following reasons. First, the super 
Green function ${\hat G}$, from which ${\hat f}$ is constructed, 
is not periodic. Rather, it satisfies
\beqa
{\hat G}({\hat\tau}_1+{\hat T},{\hat\tau}_2) 
& = & {\hat G}({\hat\tau}_1,{\hat\tau}_2) \quad \mbox{if} \quad
\tau_1 < \tau_2 \label{periodicitya} \\
{\hat G}({\hat\tau}_1-{\hat T},{\hat\tau}_2) 
& = & {\hat G}({\hat\tau}_1,{\hat\tau}_2) \quad \mbox{if} \quad
\tau_1 > \tau_2 \ , \label{periodicityb}
\eeqa
under the shift of a super period $\pm{\hat T}$ 
\beq
{\hat\tau}=(\tau,\theta) \quad\ra\quad 
{\hat\tau}\pm{\hat T}=(\tau\pm T,-\theta)\ ,
\eeq
provided that $|\tau_1-\tau_2| < T$. Secondly, we have to take 
account of the similar {\it restricted periodicity} for super derivatives 
of the super Green functions. Namely those are 'anti-periodic' if 
differentiating the shifting argument, and 'periodic' if differentiating 
the one not shifting
\beqa
D_1{\hat G}({\hat\tau}_1\pm{\hat T} ,{\hat\tau}_2) &= & 
-D_1{\hat G}({\hat\tau}_1,{\hat\tau}_2) \ , \label{D1Gshift}\\
D_2{\hat G}({\hat\tau}_1\pm{\hat T} ,{\hat\tau}_2) &=&
D_2{\hat G}({\hat\tau}_1,{\hat\tau}_2) \ , \label{D2Gshift}
\eeqa
where $\pm$ is understood as the same ordering as in \eq{periodicitya} and 
\eq{periodicityb}. Thirdly, we have to notice the following structure of 
${\hat f}$. After the Wick contractions, ${\hat f}$ becomes a polynomial 
such that every term contains all $D_i$, $i=1,\dots,n$, only {\it once} 
for each. According to this, though same arguments may appear some times, 
however the differentiated one appears exactly once for each ${\hat x}_i$. 
Therefore ${\hat f}$ behaves as if {\it anti-periodic} when one of the arguments 
is shifted by $\pm{\hat T}$ because of \eq{D1Gshift} and \eq{D2Gshift}. 
Note that ${\hat f}$ in the bosonic case behaves as if {\it periodic}.

Let us demonstrate how these things work in the case of $n=1$ (vacuum diagram). 
\beqa
{\cal A}_0& = & \int_0^T \mbox{d}{\hat x}_0 \int_{-x_0}^{-x_0+T}
\mbox{d}{\hat x}_1 \ {\hat f}({\hat 0},{\hat x}_1)  \\
& = & \int_0^T\mbox{d}{\hat x}_0\left( \int_{-x_0}^0\mbox{d}{\hat x}_1\, 
{\hat f}({\hat0},{\hat x}_1) + \int_{0}^{-x_0+T}\mbox{d}{\hat x}_1\, 
{\hat f}({\hat0},{\hat x}_1)\right)\ . \nn 
\eeqa
In the first term, $x_1$ is the smallest of the two arguments, 
since $x_1<0$ and the other is zero. 
Owing to eqs.~\eq{periodicitya}, \eq{D1Gshift} and \eq{D2Gshift} 
(for $\tau_1<\tau_2$), we can rewrite
\beq
{\hat f}({\hat0},{\hat x}_1) = - {\hat f}({\hat0},{\hat x}_1+{\hat T})\ .
\eeq
If we change
integration variables from ${\hat x}_1$ to 
${\hat x}_1'=(x_1',\theta_1') = (x_1+T,-\theta_1)$ we then obtain
\beqa
{\cal A}_0 & = & \int_0^T \mbox{d}{\hat x}_0  
\left( \int_{-x_0+T}^T \mbox{d} {\hat x}_1' {\hat f}({\hat0},{\hat x}_1') +
\int_{0}^{-x_0+T} \mbox{d} {\hat x}_1 {\hat f}({\hat0},{\hat x}_1) \right) \\
& = & \int_0^T \mbox{d}{\hat x}_0 \int_0^T
\mbox{d} {\hat x}_1 {\hat f}({\hat0},{\hat x}_1) \ = \  
T\int\theta_0 \int_0^T\mbox{d}{\hat x}_1{\hat f}({\hat 0},{\hat x}_1) \ . \nn
\eeqa

In the general cases, we just repeat the procedure as discussed in \cite{RS2}, 
and finally we conclude 
\beqa
{\cal A}_N & = & \int_0^T \mbox{d} {\hat x}_0 \int_0^T \mbox{d} {\hat x}_1 
\ldots \int_0^T \mbox{d} {\hat x}_n \ 
{\hat f}({\hat 0},{\hat x}_1,\ldots,{\hat x}_n) \\
& = & T \int d\theta_0 \int_0^T \mbox{d} {\hat x}_1 \ldots 
\int_0^T\mbox{d}{\hat x}_n\ 
{\hat f}({\hat0},{\hat x}_1,\ldots,{\hat x}_n) \ . \nn
\eeqa

%\newpage
%%%%%%%%%%%%%%%%%%%%%%%%%%%%%%%%%%%%%%%%%%%%%%%%%%%%%%%%%%%%%%%%%%%%%%%%
%                           REFERENCES                                 %
%%%%%%%%%%%%%%%%%%%%%%%%%%%%%%%%%%%%%%%%%%%%%%%%%%%%%%%%%%%%%%%%%%%%%%%%

%
\end{document}